\documentclass[pra,aps,twocolumn,superscriptaddress]{revtex4}

\usepackage{psfig}

\begin{document}

\title{Entanglement generation in continuously coupled parametric generators}

\author{Ji\v{r}\'{\i} Herec}
\affiliation{Department of Optics, Palack\'y University,
17. listopadu 50, 772 00 Olomouc, Czech Republic}

\author{Jarom\'{\i}r Fiur\'{a}\v{s}ek}
\affiliation{Department of Optics, Palack\'y University,
17. listopadu 50, 772 00 Olomouc, Czech Republic}
\affiliation{Ecole Polytechnique, CP 165/59, 
Universit\'{e} Libre de Bruxelles, 1050 Bruxelles, Belgium}

\author{Ladislav Mi\v{s}ta Jr.}

\affiliation{Department of Optics, Palack\'y University,
17. listopadu 50, 772 00 Olomouc, Czech Republic}
\date{\today}

\begin{abstract}

We investigate a compact source of entanglement. This device is composed
of a pair of linearly coupled nonlinear waveguides operating by means of
degenerate parametric downconversion. For the vacuum state at the input
the generalized squeeze variance and logarithmic negativity are used to
quantify the amount of nonclassicality and entanglement of output beams.
Squeezing and entanglement generation for various dynamical regimes of
the device are discussed.
\end{abstract}

\maketitle
\section{Introduction}

Quantum entanglement and its consequences puzzled physicists since
the famous paper by Einstein, Podolsky and Rosen \cite{Einstein35}. It
was recognized in recent years that the entanglement is not only an
intriguing and mind boggling feature of quantum mechanics but also a
crucial and extremely useful resource for information processing. The
various protocols relying on the entanglement include quantum teleportation
\cite{Bennett93}, entanglement swapping \cite{Pan98}, dense coding \cite
{Bennett92}, quantum cryptography \cite{Ekert91} and quantum computing
\cite{Barenco95}. The efficiency of quantum information processing
significantly depends on the degree of entanglement of the quantum
state shared by the parties involved in a given protocol. It is
therefore highly desirable to establish reliable sources of pure
strongly entangled states.

Quantum optics provides a natural and convenient framework for the
generation of entangled states, their manipulation and measurement.
The above-mentioned protocols have been originally
established for the quantum systems with two-dimensional Hilbert spaces
-- the qubits.  The polarization entangled pairs of photons generated by
means of the spontaneous parametric downconversion proved to be a very good
source of entangled qubits and they were employed in a large number of
experimental demonstrations of quantum information processing.

Recently, however, a significant attention has been paid to the
quantum information processing (QIP) with systems whose Hilbert space is
infinite dimensional. A typical example of such a system is a single
mode of the optical field whose Hilbert space is spanned by the infinite
(but countable) number of the Fock states $|n\rangle$.  It is in the
spirit of this latter approach that one considers the quadrature
operators $\hat{x}$, $\hat{p}$ and one speaks about QIP with continuous
variables (CV). Similarly as in the case of qubits, the parametric
downconversion provides a source of CV entanglement. Here, the relevant
entangled state is the so-called two-mode squeezed vacuum (sometimes also
called the twin beam) that can be either prepared by nondegenerate
downconversion or via mixing of two single-mode squeezed states on a
balanced beam splitter. This last-mentioned approach has been in fact used
in the experiment on the teleportation of continuous variables \cite{Furusawa98}.
The whole setup for the generation of the CV entangled state is certainly
nontrivial and quite complicated. A natural question arises whether it
could be possible to integrate the generation of the single mode squeezed
states and the mixing on a beam splitter into a single small and compact
device that would serve as a source of the entanglement.

In this paper we show that the nonlinear optical couplers that have been
thoroughly investigated during recent years both theoretically and
experimentally (see \cite{Perina00, Fiurasek01} and references therein)
can serve exactly for that purpose. We theoretically investigate
the generation of squeezing and entanglement in the nonlinear optical
coupler consisting of a pair of nonlinear waveguides operating by means
of the degenerate parametric downconversion. The waveguides are placed
close together so that the two modes can overlap and exchange energy via
linear coupling. We study the dependence of the generalized squeezing
as well as CV entanglement on the coupler's parameters. The entanglement
is quantified by the logarithmic negativity \cite{Vidal02}, which is an
easily computable entanglement monotone. Finally, the optimum phase
matching condition that yields the maximal entanglement is analyzed.
\begin{figure}[b]
\centerline{\psfig{width=0.8\linewidth,angle=0,file=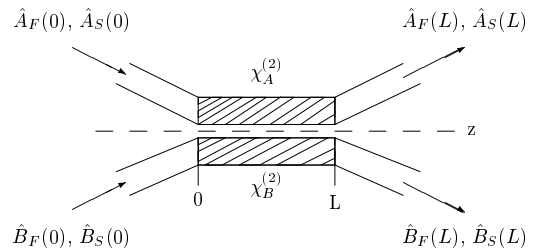}}
\caption{Sketch of nonlinear coupler formed from two nonlinear
waveguides $A$ and $B$ with susceptibilities ${\chi}^{(2)}$.}
\end{figure}
\section{The model and the equations of motion}

The device under study is schematically depicted in Fig.~1. The nonlinear
coupler consists of two linearly coupled  waveguides where the
degenerate parametric downconversion takes place. A `blue' photon with
frequency $2\omega$ may split into a pair of identical `red' photons and
also an inverse process may occur. Assuming a strong coherent pumping of
the modes with frequency $2\omega$ (second-harmonic modes) and a linear
coupling between the modes with frequency $\omega$ (fundamental modes)
in the lossless and nondispersive media, we can describe such a device
by the following effective interaction momentum operator \cite{Perina96}
\begin{equation}\label{interaction moment}
\hat{G}_{\rm int}=\hbar\left(g_{A}^{\ast}\hat{A}_{F}^2+g_{B}^{\ast}
\hat{B}_{F}^2+2g_{L}\hat{A}_{F}\hat{B}_{F}^{\dag}+\mbox{h.c.}\right)\,,
\end{equation}
where $\hat{A}_{F} (\hat{A}_{F}^{\dag})$ is annihilation (creation)
operator of the fundamental mode in the first waveguide and $\hat{B}_{F}
(\hat{B}_{F}^{\dag})$ is annihilation (creation) operator of the
fundamental mode in the second waveguide. The constant $g_{A} (g_{B})$ is
a product of the strong coherent amplitude of the second-harmonic
mode in the first (second) waveguide and the nonlinear coupling
constant, which is proportional to the quadratic susceptibility
${\chi}^{(2)}_A$ (${\chi}^{(2)}_B$). The constants $g_A$ and $g_B$
determine the efficiency of the downconversion processes.
Linear coupling constant denoted as $g_L$ describes the energy exchange
between the waveguides by means of coupling via the evanescent waves.
The symbol $\hbar$ denotes the reduced Planck constant and h.c.
represents Hermitian conjugate terms.

We shall work in the interaction picture where the spatial evolution of
an operator $\hat{O}$ is governed by the Heisenberg-like equation (see
\cite{Perina00, Fiurasek01} and references therein)
\begin{equation}
i\hbar\frac{d\hat{O}}{dz}=[\hat{G}_{\rm int},\hat{O}]\,.
\end{equation}
Here $z$ is a spatial coordinate along the direction of propagation,
$\hat{G}_{\rm int}$ is an interaction momentum operator and $[,]$
denotes a commutator. Using the momentum operator (\ref{interaction
moment}), we obtain the Heisenberg equations of motion in the
interaction picture of the form:
\begin{eqnarray}\label{equation of motion}
\frac{d\hat{A}_{F}}{dz}&=&2ig_{A}\hat{A}_{F}^{\dag}+2ig_{L}^{\ast}
\hat{B}_{F}\,,\nonumber\\
\frac{d\hat{B}_{F}}{dz}&=&2ig_{B}\hat{B}_{F}^{\dag}+2ig_{L}\hat{A}_{F}\,.
\end{eqnarray}
For our purposes it is convenient to deal with the quadrature
components $\hat{x}$, $\hat{p}$ rather than with the creation and
annihilation operators. The quadratures can be directly measured via
balanced homodyne detection.
Recall that the quadrature components can be expressed as linear
combinations of the creation and annihilation operators as follows:
\begin{eqnarray}\label{quadrature operators}
\hat{x}_{A}&=&\frac{\sqrt 2}{2}\left(\hat{A}_{F}^{\dag}+\hat{A}_{F}
\right)\,,\quad
\hat{p}_{A}=i\frac{\sqrt 2}{2}\left(\hat{A}_{F}^{\dag}-\hat{A}_{F}
\right)\,,\nonumber\\
\hat{x}_{B}&=&\frac{\sqrt 2}{2}\left(\hat{B}_{F}^{\dag}+\hat{B}_{F}
\right)\,,\quad
\hat{p}_{B}=i\frac{\sqrt 2}{2}\left(\hat{B}_{F}^{\dag}-\hat{B}_{F}
\right)\,.\nonumber\\
\end{eqnarray}
We further express the (generally complex) coupling constants in
terms of their amplitudes and phases,
\begin{equation}\label{decomposition}
g_{j}=|g_{j}|\mbox{exp($i\varphi_{j}$)}\quad\mbox{for}\quad j=L,A,B\,.
\end{equation}
With these definitions at hand, we can derive from Eq.~(\ref{equation
of motion}) the Heisenberg equations for the quadrature operators. We
write them down in the compact matrix form,
\begin{equation}\label{matrix equation of motion}
\frac{d\hat{\xi}}{dz}={\bf M}\hat{\xi}\,,
\end{equation}
where
\begin{equation}\label{matrices}
{\bf M}=\left(\begin{array}{cccc}
-S_{A} & C_{A} & S_{L} &-C_{L}\\
C_{A} & S_{A} & C_{L} & S_{L}\\
-S_{L} & -C_{L} & -S_{B} & C_{B}\\
C_{L} & -S_{L} & C_{B} & S_{B}\\
\end{array}
\right)\,,\quad
\hat{\xi}=\left(\begin{array}{c}
\hat{x}_{A}\\
\hat{p}_{A}\\
\hat{x}_{B}\\
\hat{p}_{B}\\
\end{array}
\right)\,,
\end{equation}
and
\begin{equation}\label{functions S and C}
S_{j}=2|g_{j}|\sin\left(\varphi_{j}\right)\,,\qquad
C_{j}=2|g_{j}|\cos\left(\varphi_{j}\right)\,.
\end{equation}
Eq.~(\ref{matrix equation of motion}) represents a system of linear
differential equations with constant coefficients that  can be solved
very simply by determining the eigenvalues and eigenvectors of the
matrix ${\bf M}$. Explicit analytical expressions for the solution
can be found in \cite{Perina96}. Here we write the solution in a formal
matrix notation
\begin{equation}\label{solution of zero mismatch}
\hat{\xi}(z)={\bf S}(z)\hat{\xi}(0)\,,\qquad {\bf S}(z)=e^{{\bf M}z}\,.
\end{equation}
This evolution represents a linear canonical transformation of the
quadrature operators. All possible linear canonical transformations
of $N$ modes form a group of symplectic (also called Bogolyubov)
transformations $Sp(2N,R)$. In our case, for each $z$ the matrix
${\bf S}(z)$ is an element from the four dimensional representation
of the symplectic group $Sp(4,R)$.

One of the most important features of the
symplectic transformations is that they transform Gaussian states again
onto Gaussian states. The Gaussian states are those whose Wigner
function is a Gaussian. Such states are thus fully characterized by the
first and second moments of the quadrature operators. Moreover, the
first moments (mean values) of the quadratures can be always set to zero
via appropriate  displacements applied locally to each mode. Since we
are interested in entanglement properties of the state generated in the
coupler, we can focus solely on the evolution of the second moments.
\section{The generalized squeeze variance}

It is convenient to arrange the second moments into the covariance
matrix ${\bf V}$ whose elements are defined as follows \cite{Simon94}:
\begin{equation}\label{covariance matrix}
V_{jk}=\frac{1}{2}\left(\langle\Delta\hat{\xi}_{j}\Delta\hat{\xi}_{k}
\rangle+\langle\Delta\hat{\xi}_{k}\Delta\hat{\xi}_{j}\rangle\right)\,.
\end{equation}
Here $\Delta\hat{\xi}_{j}=\hat{\xi}_{j}-\langle\hat{\xi}_{j}\rangle$
and $\langle\hat{\xi}_{j}\rangle=\mbox{Tr}(\hat{\rho}\hat{\xi}_{j})$.
In our analysis of the coupler we shall assume that all processes are
spontaneous, i.e., both fundamental modes are initially in the vacuum
state. The covariance matrix corresponding to this input state is
proportional to the identity matrix,
\begin{equation}\label{input state}
{\bf V}(0)=\frac{1}{2}\left(
\begin{array}{cccc}
1 & 0 & 0 & 0\\
0 & 1 & 0 & 0\\
0 & 0 & 1 & 0\\
0 & 0 & 0 & 1
\end{array}
\right)\,.
\end{equation}
The evolution of the covariance matrix corresponding to the symplectic
transformation (\ref{solution of zero mismatch}) is given by the
following formula:
\begin{equation}\label{evolution of covariance matrix}
{\bf V}(z)={\bf S}(z){\bf V}(0){\bf S}^{T}(z)\,,
\end{equation}
where $T$ stands for the transposition.

Quantum entanglement is one of the manifestations of the
nonclassical features of the quantum states. The CV quantum state is
said to have a classical analogue if its density matrix can be written
as a convex mixture of projectors on coherent states. Such states cannot
be entangled and they can exhibit only classical correlations. Since the
nonclassicality is a necessary prerequisite for the entanglement \cite{
Buzek02} it is of interest to investigate whether and how much nonclassical
are the states generated in the nonlinear coupler. This issue has been
addressed in many recent papers devoted to the analysis of the
quantum-statistical properties of light propagating in
nonlinear couplers \cite{Perina00, Fiurasek01}. Various kinds of
nonclassical behaviour have been predicted, such as the generation of
light with sub-Poissonian photon number statistics, and single-mode or
two-mode squeezed light.

For Gaussian states, the nonclassicality criterion
is particularly simple because these states are nonclassical if and only
if (iff) they are squeezed. More formally, we say that the (generally $N$-mode)
Gaussian state is nonclassical iff there exists a quadrature $\hat{X}$
that is a linear combination of the quadratures $\hat{x}_j$ and $\hat{p}_j$
of the $N$ involved modes such that the variance of $\hat{X}$ is below
the coherent-state level $1/2$. The smallest variance obtained as
minimum over the variances of all possible linear combinations of the
quadratures $\hat{x}_j$ and $\hat{p}_j$ is called the generalized squeeze
variance $\lambda$ \cite{Simon94}. This variance can be calculated as the
lowest eigenvalue of the covariance matrix (\ref{evolution of covariance
matrix})
\begin{equation}
\lambda(z)={\rm min}\{{\rm eig}[{\bf V}(z)]\}\,.
\end{equation}
The Gaussian state is nonclassical and squeezed iff $\lambda<1/2$.
\begin{figure}[t]
\centerline{\psfig{width=0.8\linewidth,angle=0,file=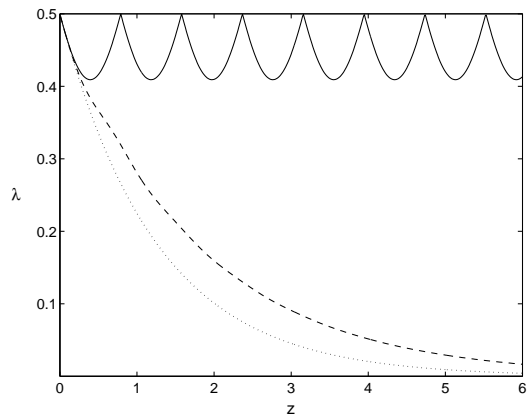}}
\caption{The dependence of the generalized squeeze variance $\lambda$
on the length $z$ of the symmetric coupler operating below the
threshold, $|g_L|=2$, $|g_A|=|g_B|=0.2$. The three curves correspond to
the phase differences $\Delta\phi=0$ (solid line),
$\Delta\phi=\pi/2$ (dashed line), and $\Delta\phi=\pi$ (dotted line).}
\end{figure}
In our discussion of the coupler's behavior we distinguish two main
regimes of operation. In the first regime, the linear coupling dominates
over the nonlinear interactions
\begin{equation}\label{threshold condition}
2|g_L|>|g_A+g_B|
\end{equation}
and the coupler operates below the threshold. If the opposite inequality
holds then the nonlinear interactions in the waveguides become dominant
and the coupler operates above the threshold. The threshold condition
($2|g_L|=|g_A+g_B|$) was determined by analyzing the eigenvalues of the
matrix ${\bf M}$ (see also Ref.~\cite{Fiurasek00}). These eigenvalues
are complex below the threshold, which leads to oscillatory dynamics
typical for the linear coupling. Above the threshold, all eigenvalues
are real and the dynamics of the coupler is reminiscent of a pure
amplification process.

The behavior of the coupler also significantly depends on the phase
matching condition. It can be inferred from the analytical expressions
for the symplectic transformation ${\bf S}(z)$ that there is a single
effective phase difference
\begin{equation}\label{phase difference}
\Delta\phi=\phi_A-\phi_B+2\phi_L
\end{equation}
that controls the dynamics of the coupler.

In figure 2 we plot the dependence of the generalized squeeze variance
$\lambda$ on the length $z$ of the symmetric coupler ($|g_A|=|g_B|$) that
operates below the threshold. The three curves correspond to three
different choices of  the phase difference $\Delta\phi$. For
$\Delta\phi=0$ we observe a periodic evolution of $\lambda(z)$.
In this specific case, the eigenvalues of ${\bf M}$ are purely imaginary,
the dynamics is purely oscillatory, and the quantum state of the two
modes periodically returns to the initial vacuum state. For the two
other choices $\Delta\phi=\pi/2$ and $\Delta\phi=\pi$ the parametric
amplification  sets on and the squeeze variance exponentially
decreases with growing length of the coupler. The squeezing is
fastest for $\Delta\phi=\pi$. For comparison, we plot in figure 3 the
function $\lambda(z)$ for symmetric coupler operating above the threshold.
It follows that in this case the phase difference almost does not
influence the dynamics of squeezing and the three curves shown in Fig.~3
corresponding to three different phase shifts $\Delta\phi$ almost coincide.

For the asymmetric coupler ($|g_A|\not =|g_B|$) below the threshold, the
eigenvalues of ${\bf M}$ have nonzero both imaginary and real parts giving
rise to periodically modulated exponential decrease of $\lambda(z)$. The
modulations smooth out as $\Delta\phi$ goes to $\pi$. Above the
threshold $\lambda(z)$ exhibits qualitatively the same behavior as
in the symmetric case.
\begin{figure}[t]
\centerline{\psfig{width=0.8\linewidth,angle=0,file=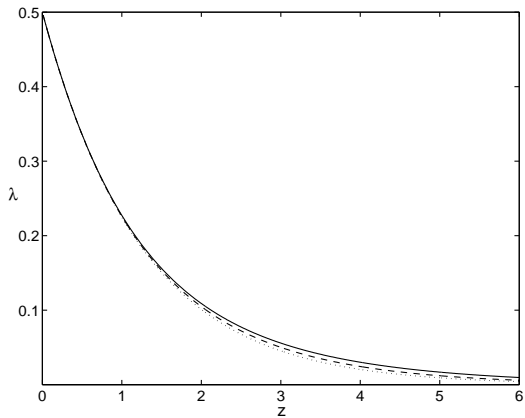}}
\caption{The dependence of the generalized squeeze variance $\lambda$
on the length $z$ of the symmetric coupler operating above the
threshold, $|g_L|=0.15$, $|g_A|=|g_B|=0.2$. The three curves correspond to
the phase differences $\Delta\phi=0$ (solid line),
$\Delta\phi=\pi/2$ (dashed line), and $\Delta\phi=\pi$ (dotted line).}
\end{figure}

\section{Separability and entanglement}

Having discussed the squeezing of light in the coupler, we now turn our
attention to the generation of entangled states. A state $\hat{\rho}_{AB}$
of two subsystems $A$ and $B$ is entangled iff $\hat{\rho}_{AB}$ cannot
be written as a convex mixture of product states
\begin{equation}\label{bipartite entanglement state}
\hat{\rho}_{AB} \neq \sum_j p_j \hat{\rho}_{A,j}\otimes \hat{\rho}_{B,j}\,,
\qquad p_j>0,
\end{equation}
where $\hat{\rho}_{A,j}$ and $\hat{\rho}_{B,j}$ are states of subsystems
$A$ and $B$. Since we assume that the fundamental modes are initially in pure
vacuum states and the evolution is unitary, the two-mode state is pure for all
$z$. It holds that every pure bipartite state $|\psi\rangle_{AB}$ that is
not a product state is entangled. While it is easy to decide whether a
pure state is separable or not, this problem is much more difficult for
the general mixed bipartite state. Several separability criteria have
been proposed in the literature. Among them, the most powerful and
important is the Peres-Horodecki (PH) criterion \cite{Horodecki95,
Peres96, Horodecki96}. According to this criterion, a quantum state
$\hat{\rho}_{AB}$ is entangled if the partially transposed density matrix
\begin{equation}
(\hat{\rho}_{AB})_{m\alpha,n\beta}^{T_{A}}\equiv(\hat{\rho}_{AB})_
{n\alpha,m\beta}
\end{equation}
is nonpositive, i.e., it has at least one negative
eigenvalue. In general, this is only a sufficient condition on
entanglement and there exist entangled states with positive partially
transposed density matrix (the so-called bound entangled states).
However, in case of two-mode bipartite Gaussian states this criterion
turns out to be both sufficient and necessary \cite{Duan00, Simon00}.
\begin{figure}[t]
\centerline{\psfig{width=0.8\linewidth,angle=0,file=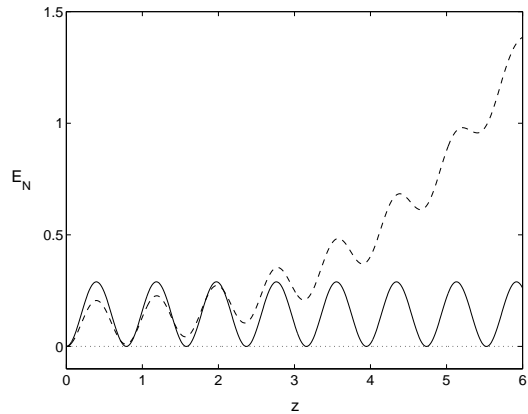}}
\caption{The dependence of the logarithmic negativity $E_N$ on the length
$z$ of the symmetric coupler operating below the threshold, $|g_L|=2$,
$|g_A|=|g_B|=0.2$. The three curves correspond to the phase differences
$\Delta\phi=0$ (solid line), $\Delta\phi=\pi/2$ (dashed line), and
$\Delta\phi=\pi$ (dotted line).}
\end{figure}
The PH criterion is qualitative, it tells us whether the state is
entangled or separable, but it does not quantify the amount of the
entanglement present in that state. For this purpose, various
entanglement measures have been proposed. The PH criterion suggests that
a function of the negative eigenvalues of the partially transposed
density matrix could be a good entanglement measure. Indeed, such
measure can be constructed and it is called the
logarithmic negativity \cite{Vidal02}
\begin{equation}\label{logarithmic negativity}
E_N(\hat{\rho})=\log_2 [1+2{\cal{N}}(\hat{\rho})]\,.
\end{equation}
Here the negativity ${\cal{N}}(\hat{\rho})$ is the 
sum of the absolute values of the negative eigenvalues $\mu_j$ 
of the partially transposed matrix $\hat{\rho}^{T_A}_{AB}$
\begin{equation}\label{negativity}
{\cal{N}}(\hat{\rho})=\sum_j |\mu_j|\,.
\end{equation}
The logarithmic negativity $E_N$ has the great advantage that it can be
easily computed for any bipartite Gaussian state. In fact, analytical
formula for $E_N$ has been derived for the general mixed two-mode
Gaussian state. The details can be found in \cite{Vidal02}, here we only
mention the main results and formulas.

It is convenient to decompose the covariance matrix as follows
\begin{equation}\label{decomposition1}
{\bf V}=\left(
\begin{array}{cc}
{\bf A} & {\bf C}\\
{\bf C}^T & {\bf B}\\
\end{array}
\right)\,,
\end{equation}
where ${\bf A}$ and ${\bf B}$ are the covariance matrices of the modes
on Alice's and Bob's sides, respectively, and ${\bf C}$ contains the
intermodal correlations. We must calculate the symplectic
spectrum $(c_1,c_2)$ of ${\bf V}$  that can be obtained by solving the
biquadratic equation (only positive roots are considered)
\begin{equation}\label{biquadratic equation}
\zeta^4+\left(\mbox{det}{\bf A}+\mbox{det}{\bf B}-2\mbox{det}
{\bf C}\right)\zeta^2+\mbox{det}{\bf V}=0\,.
\end{equation}
The logarithmic negativity is a function of the symplectic eigenvalues
\begin{equation}\label{function of symplectic spectrum}
E_N=\sum_{j=1}^{2} F(c_{j})\,,
\end{equation}
where the function $F(c)$ reads
\begin{equation}\label{symplectic spectrum}
F(c)=\left\{
\begin{array}{cc}
0 & \mbox{for}\, 2c\geq 1\\
-\log_2(2c) & \mbox{for}\, 2c<1\,.
\end{array}\right.
\end{equation}
\begin{figure}[t]
\centerline{\psfig{width=0.8\linewidth,angle=0,file=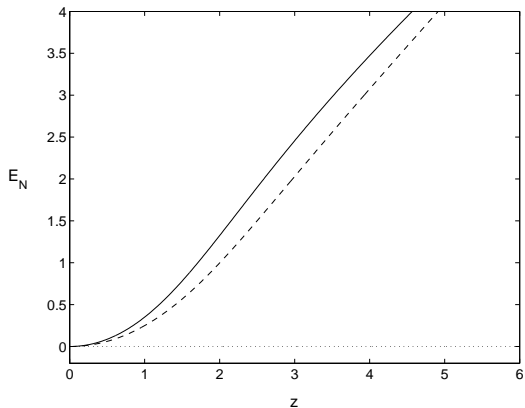}}
\caption{The dependence of the logarithmic negativity $E_N$ on the length
$z$ of the symmetric coupler operating above the threshold, $|g_L|=0.15$,
$|g_A|=|g_B|=0.2$. The three curves correspond to the phase differences
$\Delta\phi=0$ (solid line), $\Delta\phi=\pi/2$ (dashed line), and
$\Delta\phi=\pi$ (dotted line).}
\end{figure}
The evolution of the logarithmic negativity is plotted in Fig.~4. The
parameters are the same as in Fig.~2, so the coupler is below the
threshold. As could have been expected, for $\Delta\phi=0$ we get a
purely oscillatory evolution and $E_N$ periodically attains its maximum.
The oscillations are clearly present also for $\Delta\phi=\pi/2$ but
the general trend is that $E_N$ grows with $z$. This is in agreement
with the fact that for this detuning the state becomes more and more
squeezed with growing $z$, cf. Fig.~2. Remarkably, $E_N(z)=0$ for all
$z$ when $\Delta\phi=\pi$. For this particular phase difference, the
coupler behaves like two decoupled single mode squeezers that do not
interact at all \cite{Fiurasek00}. Figure 5 shows the behavior of $E_N(z)$
when the coupler operates above the threshold. We can see that the
oscillatory evolution is replaced with a monotonic growth of $E_N(z)$
with $z$. However, the amount of entanglement significantly depends on
$\Delta\phi$ even when the coupler is above the threshold. This is in
marked difference with the generalized squeeze variance that is to a
large extent insensitive to the choice of $\Delta\phi$, see Fig.~3.

In the case of the asymmetric coupler above the threshold $E_N$
grows monotonously with growing $z$. For small $z$ the rapidity of
the growth decreases as $\Delta\phi$ goes from $0$ to $\pi$. Below
the threshold $E_N$ behaves similarly as in the symmetric case. The
only difference occurs for $\Delta\phi=\pi$ when $E_N$ changes
periodically and no decoupling of the waveguides takes place.
\begin{figure}[t]
\centerline{\psfig{width=0.8\linewidth,angle=0,file=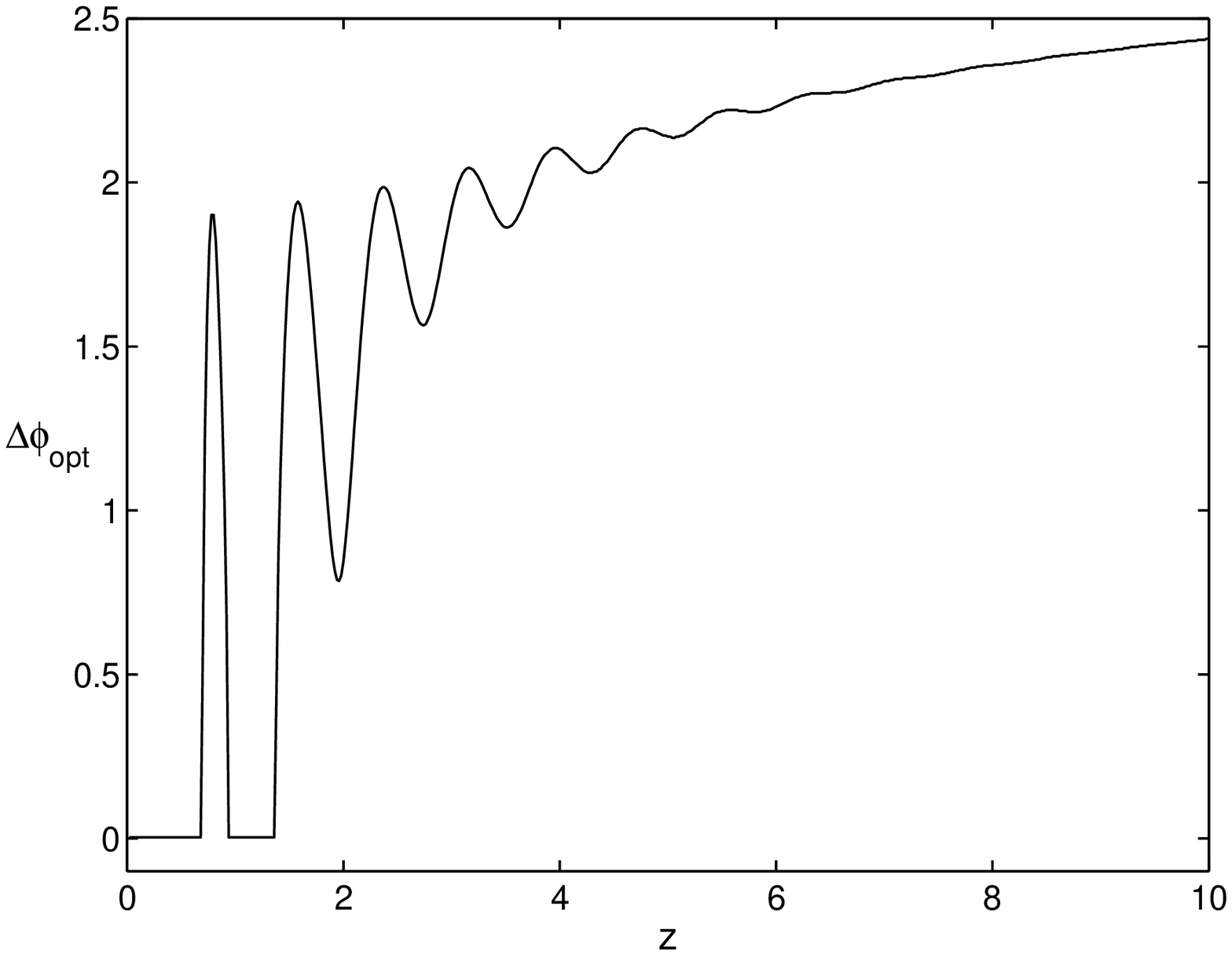}}
\centerline{\psfig{width=0.8\linewidth,angle=0,file=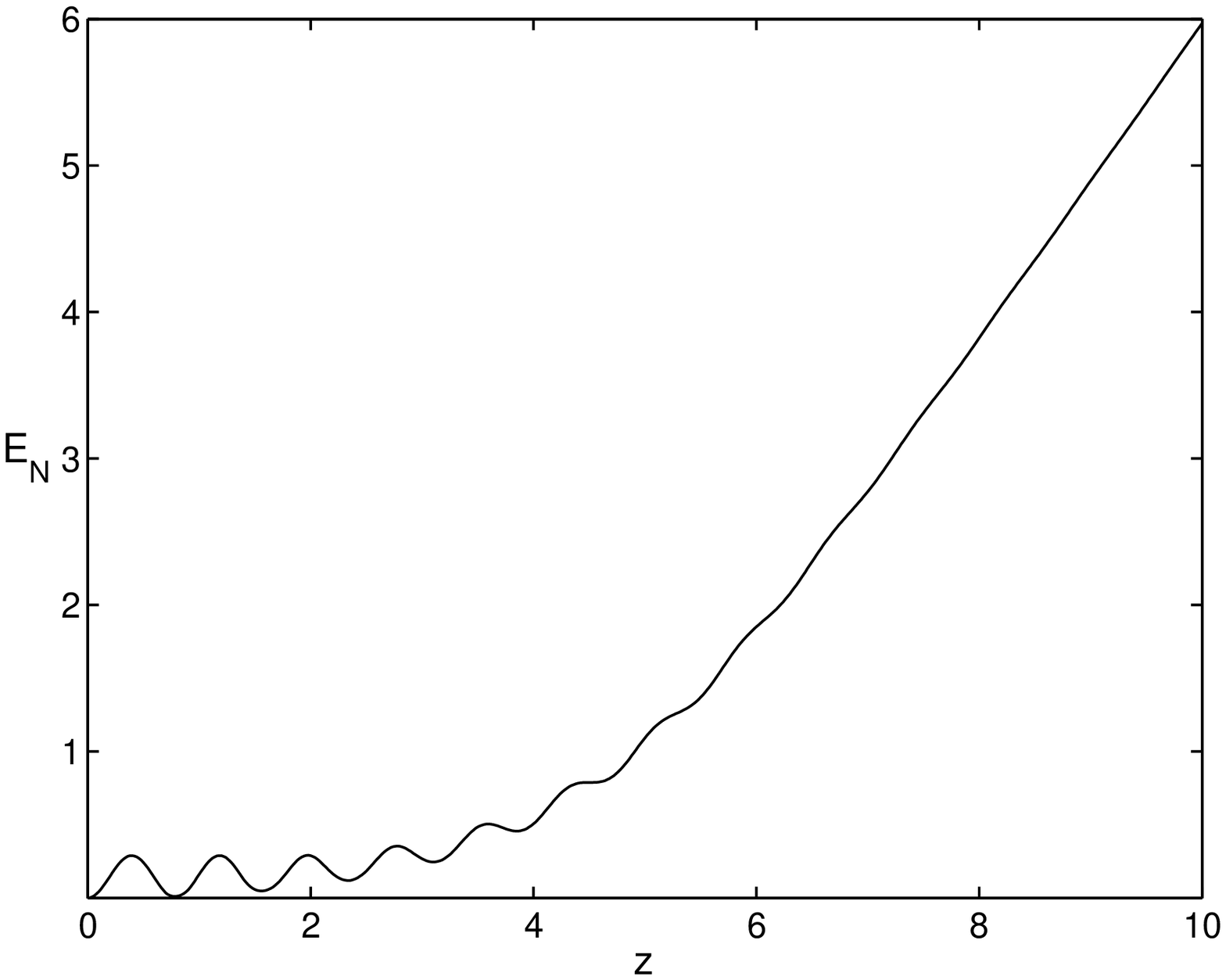}}
\caption{The dependence of the optimal phase difference
$\Delta\phi_{opt}$ and the maximum of the logarithmic negativity
$E_N$ on the length $z$ of the symmetric coupler operating below
the threshold $|g_L|=2$, $|g_A|=|g_B|=0.2$.}
\end{figure}

The results depicted in Figs.~4 and 5 indicate that for each $z$ and a
given set of parameters, we may tune the phase difference such as to
maximize the entanglement. We have performed numerical calculations and
determined the optimal phase difference $\Delta\phi_{\rm opt}$ and the
corresponding maximal achievable $E_N$. The results for the coupler
operating below the threshold are presented in Fig.~6. The
$\Delta\phi_{\rm opt}$ exhibits an interesting behavior. Initially, it is
optimal to set $\Delta\phi_{\rm opt}=0$ and this is optimal up to certain length
$z_0$. At this distance, the optimal phase difference becomes to
deviate from $0$ and $\Delta\phi_{\rm opt}$ oscillates with $z$. The
figure 6 suggests that for large $z$, $\Delta\phi_{\rm opt}$ approaches some
fixed asymptotic value. We note that similar results were obtained also
for the coupler above the threshold.
\section{Conclusions}

In this paper we have explored the controllable compact source of
squeezing and entanglement formed by a nonlinear optical coupler
composed of two nonlinear waveguides operating by means of spontaneous
degenerate parametric downconversion. Firstly, the behavior of the
generalized squeeze variance in the below the threshold and above
the threshold regimes has been analyzed. Secondly, by calculating
the logarithmic negativity the possibility of the entanglement
generation in the coupler has been demonstrated. In particular, it was
shown that the amount of the entanglement can be controlled via the
effective phase difference (\ref{phase difference}). Finally, with the
help of the numerical calculation the optimal effective phase difference
providing maximum entanglement at the output of the coupler has been
found. Our theoretical analysis thus clearly illustrates the potential
utility of the coupler for controlled generation of CV entanglement.
\section{Acknowledgments}

This work was supported by Project LN00A015 and Research Project No.
CEZ: J14/98 of the Czech Ministry of Education and by the EU grant
under QIPC Project IST-1999-13071 (QUICOV).

\end{document}